\documentclass[twocolumn,canadian]{aastex6}
\usepackage[T1]{fontenc}
\usepackage[latin9]{inputenc}
\usepackage{verbatim}
\usepackage{graphicx}

\makeatletter
\slugcomment{}
\shorttitle{}
\shortauthors{}

\makeatother

\usepackage{babel}
\begin{document}

\title{3D visualization of astronomy data cubes using immersive displays}

\author{Gilles Ferrand}

\affil{Dept. of Physics and Astronomy, University of Manitoba}

\email{gferrand@physics.umanitoba.ca}

\author{Jayanne English}

\affil{Dept. of Physics and Astronomy, University of Manitoba}

\author{Pourang Irani}

\affil{Human-Computer Interaction Lab, Dept. of Computer Science, University
of Manitoba}
\begin{abstract}
We report on an exploratory project aimed at performing immersive
3D visualization of astronomical data, starting with spectral-line
radio data cubes from galaxies. This work is done as a collaboration
between the Department of Physics and Astronomy and the Department
of Computer Science at the University of Manitoba. We are building
our prototype using the 3D engine Unity, because of its ease of use
for integration with advanced displays such as a CAVE environment,
a zSpace tabletop, or virtual reality headsets. We address general
issues regarding 3D visualization, such as: load and convert astronomy
data, perform volume rendering on the GPU, and produce physically
meaningful visualizations using principles of visual literacy. We
discuss some challenges to be met when designing a user interface
that allows us to take advantage of this new way of exploring data.
We hope to lay the foundations for an innovative framework useful
for all astronomers who use spectral line data cubes, and encourage
interested parties to join our efforts. This pilot project addresses
the challenges presented by frontier astronomy experiments, such as
the Square Kilometre Array and its precursors.
\end{abstract}


\section{Motivation}

One of the major challenges faced by astronomers is to digest the
large amount of diverse data generated by modern instruments or simulations.
To truly exploit the data, it is necessary to develop visualization
tools that allow exploration of all their complexity and dimensions,
an aspect that is unfortunately often overlooked. Although astronomical
data is commonly obtained in projection, producing 2-dimensional images,
the addition of spectral information can make the data 3-dimensional.
Examples include observations from Integral Field Units in the optical
regime as well as microwave and radio telescope receivers, we use
the latter for illustration below. Manipulating the resulting 3D data
cubes in a meaningful way is a non-trivial task, that requires a knowledge
of both the physics at play, and the visualization techniques involved.
We have started an interdisciplinary project at the University of
Manitoba, a collaboration of astrophysicists and computer scientists,
to investigate the use of Virtual Reality (VR) environments.

\paragraph{Radio data cubes of galaxies}

An important astrophysical process is the emission by neutral hydrogen
(HI) of a line with a rest-frame wavelength of 21 cm, that is detected
with radio telescopes. Observations of this specific emission line
are made at different wavelengths, in different receiver channels,
that correspond to the same line but shifted by the Doppler effect,
because of the motion of the emitting material \textendash{} hence
channels can be labelled as a velocity dimension. Another similar
example is the microwave line emission of carbon monoxide gas (CO).
If the source is spatially resolved, the resulting product is a 3D
data cube, that has 2~spatial dimensions and 1~velocity dimension. 

For our experimentations we have chosen the galaxy NGC~3198, a type
SB(rs)c spiral located at 9.5~Mpc. The object was observed in optical,
infrared, ultraviolet, as well as in radio. We are using data from
the THINGS survey\footnote{\url{http://www.mpia.de/THINGS/Overview.html}}
obtained at the NRAO Very Large Array (VLA). The data cube is 1024$\times$1024
pixels by 72 velocity channels, of the order of 75~million points.\footnote{Astronomical data can of course be much bigger than this, we defer
the handling of larger-than-memory data to future work, on this topic
see \citet{Hassan2011b}.}

A~typical desktop of a radio-astronomer is shown on figure~\ref{fig:Karma}.
The Karma software suite\footnote{\url{http://www.atnf.csiro.au/computing/software/karma/}}
is still widely used despite being 20 years old and no longer actively
maintained. The plots shown are a projection and a slice in the cube,
so reducing the dimensionality of the data. In this paper, we intend
on visualizing the entire data set. The Karma software does 3D volume
rendering, and lets the user define the colour transfer functions
in a precise way, but it offers very limited interactivity given its
age. The next generation tool is the Viewer application from CASA:
the Common Astronomy Software Applications package, however it does
not currently support 3D rendering, which was not identified as a
priority.\footnote{See a progress report at \url{https://science.nrao.edu/facilities/alma/alma-development-2015/VisualizationPortal.pdf},
the current focus is on porting the software to the cloud, see \citet{Rosolowsky2015b}.} Other specialized software like GAIA or even ds9 offer some 3D modes.\footnote{Another approach is to use generic 3D visualization software (see
a review of some options for radio astronomers in \citealt{Punzo2015a}),
or to write custom programs using visualization libraries (e.g. S2PLOT
by \citealt{Barnes2006a}).} However all these software are made for the desktop, and none offers
support for advanced displays. 

\paragraph{The role and challenges of 3D in scientific discovery }

In this paper we are interested in displaying the 3D data in actual
3D space, to get a holistic view, in the expectation that this will
generate a more correct perception, and help build an intuition, of
the data. We think this is important for quick interpretation, and
also necessary to discover structures that were not anticipated \textendash{}
we emphasize that our data are from observations, so they are poorly
structured and their actual content is not known in advance. The next-generation
radio facilities being developed, such as the SKA, will produce amounts
of data that will require much progress not only in terms of hardware
but also in terms of software and assembling the visualization pipeline.
While it is anticipated that automated analysis systems will be put
in place, the direct inspection of the data will still remain critical
to ensure proper operations and to foster discovery \citep{Hassan2011a}.
The human brain is wired to analyze 3D environments, and we do have
3D displays, so it seems natural to use them to visualize our 3D data.
In this regard, the interface between the machine and the human brain
is the bottleneck in the interpretation of complex astronomical data:
it was already pointed out by \citet{Norris1994a} that visualization
tools have to be more user-friendly. These days there is (again) a
lot of hope and momentum in the field of Virtual Reality (as well
as in the field of Augmented Reality, or combinations thereof).%
It has already many professional applications, in the fields of engineering,
architecture, marketing, training, health, and scientific visualization.
But developing interfaces allowing for Natural User Interaction (NUI)
in 3D is still an active field of research \textendash{} we note that
in their review of solutions for astronomers, \citet{Punzo2015a}
deliberately did not consider the new generation of cheaper 3D hardware
(such as the Leap Motion or the Oculus Rift), because of their still
uncertain fate, and also because of the lack of expertise for these
new interfaces. We think that astronomers should embrace this new
technology, and develop the interfaces they need to take advantage
of it. Producing tools and techniques that support and enhance our
research is important, so that astronomy remains at the forefront
of the field of visualization. 

\section{Tools and approach}

With our project we are moving away from the visual arts tradition
of representing the world on a canvas \textendash something we all
became acquainted with, but really is a construction and requires
training, closer to the way we actually perceive the world with our
senses. 

\paragraph{Virtual Reality displays}

On the flat display of a desktop or mobile computer, the visualization
is limited to 2D views: slices and projections, that have to be flipped
through, or a fake 3D view, emulated with tricks like perspective
and shading. Stereoscopic 3D can be achieved using dual projectors,
that present a slightly different image to each eye \textendash{}
a technique that most of us have experienced in movie theatres. The
3D displays we are considering here bring something more: the tracking
of the viewer (commonly using infrared cameras), which makes the experience
distinctively different. First this enables motion parallax, which
gives a much stronger depth cue, and second this allows direct interaction
with what is being displayed. Depending on the hardware used, one
can get the feeling of being fully immersed inside the data cube,
as if it was a physical object that we can explore and manipulate. 

3D displays for Virtual Reality come broadly into two categories:
``fish tanks'', systems where the user is looking at a fixed screen
or set of screens that define the boundaries of a virtual volumetric
screen, and head-mounted displays (HMD), systems where the user is
wearing a pair of screens attached directly in front of their eyes. 

In the first category, of ``fish tanks'', we have been experimenting
with a CAVE = Cave Automatic Virtual Environment%
by Visbox (see figure \ref{fig:visitorCAVE}) and with the zSpace%
tabletop. Both are made from flat screens, operating at HD resolution
of 1920$\times$1080 pixels. The CAVE uses dual projectors for each
screen (we have two: one wall and one floor, making 1920$\times$1080$\times$1080
voxels), while the zSpace uses a LCD screen at double the standard
refresh rate (120~Hz). Both rely on the polarization of light to
separate the left and right images, using passive polarized glasses.
Both make use of IR cameras to track the position of the glasses and
other interaction devices: the Flystick ``wand'' in the CAVE, the
built-in stylus on the zSpace. The CAVE is a human-scale device, of
several cubic meters, while the zSpace fits on a desk, offering a
smaller but also crisper view. We observed that some scientists prefer
the former, because it allows them to be immersed inside the data
cube, while others prefer the latter, because it allows them to play
with the data cube in their hands in a simple setup. 

In the second category, of HMDs, we plan to experiment with two headsets
that were very recently released for the general public: the Rift
by Oculus, and the Vive by HTC and Valve.%
These promise full immersion, and are available at a much lower price
point \textendash{} so it is important for scientists too, to see
if they will lead to mass adoption. A drawback of the headset approach
is however that the pixel resolution looks coarser, and that it is
not easy for people wanting to collaborate to share the same physical
space (for a comparison of a high-end and a low-cost solution, see
\citealt{Fluke2016a}).

\paragraph{Development with the Unity engine}

We see that there are different technical solutions available for
immersive displays, and that they rely on more hardware parts than
a standard display. Fortunately, we have reached a point in time when
it is no longer necessary to know about all the technical aspects
to harness these systems. Rather than expanding an existing scientific
visualization software to advanced displays, we took a radically different
approach, of customizing a generic 3D platform for our needs. We have
been building our prototype using Unity, the most popular engine for
game development.\footnote{\url{https://unity3d.com} \\
We have been using the (free) personal edition of Unity version~5. } While the choice of a game engine may sound surprising at first,
building on a market standard offers a lot of advantages, in particular
for fast prototyping and testing. Unity has millions of users worldwide,
and benefits from continuous development, on a scale that is normally
not accessible to scientists. It has already been used for a number
of ``serious applications'', in particular in the medical and architectural
fields, with also several experiments in the natural sciences: biology,
geology, meteorology.\footnote{The only attempt to use Unity in astrophysics, that we are aware of,
was made by a team at Caltech \citep{Cioc2013a}, that was also investigating
virtual worlds, with a focus on the visualization of multi-dimensional
data \textendash{} using the highest available number of physical
dimensions. } Locally, the Human-Computer Interaction lab has adopted it for the
development of immersive environments to simulate the devices of the
future. 

\begin{figure*}
\noindent \centering{}\includegraphics[height=0.4\textheight]{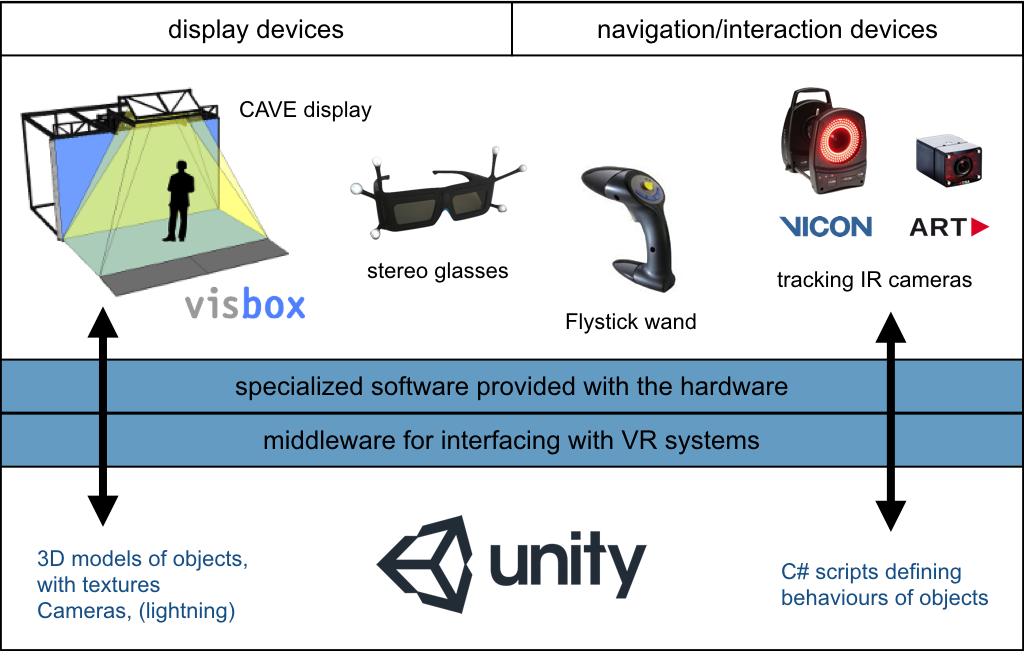}\caption{\label{fig:Unity2CAVE}Schematic illustrating how we use the Unity
software to drive the different hardware components that make the
CAVE an immersive environment. We take advantage of the visual development
interface and the software abstraction layers to design the user experience.}
\end{figure*}

Unity allows a high-level programming and designing so that we can
focus on the content. It allows for visual editing and immediate testing
of the scene and the code, which when ready can be exported as a standalone
executable. It is cross-platform, and targets all devices, including
all the advanced displays that currently exist. Support may be built-in,
or enabled via \emph{plugins} provided by hardware vendors or by third-party
specialists. See figure~\ref{fig:Unity2CAVE} for a schematics of
the software-hardware interface in the case of the CAVE. In the CAVE
we use the middleVR \emph{middleware} to interface with all the display
and interaction components; on the zSpace we use the zCore plugin
provided; all the popular headsets are also supported. This way we
could develop a demo on the desktop, and port it with minimal effort
to various VR displays, without having to worry about the handling
of multiple cameras to get the stereoscopy, or knowing about the different
drivers needed for the head tracking. This is important because scientists
need continuity in their workflow when working with different visualization
platforms. 

A~drawback of using a generic solution is that it is not tailored
for our particular needs, and so may not offer the best possible performance
for a given task. 

\section{Work accomplished}

In a period of a few months, we developed a prototype tool that loads
a radio data cube of the galaxy and renders it in 3D, on the desktop
and in VR displays, so far the CAVE and the zSpace.\footnote{We also exported to the web using WebGL, but this no longer seemed
feasible when we started to do volume rendering.} It offers basic interaction capabilities to manipulate the cube:
translate/rotate along/around any axis, scale up or down, and slice,
with whatever input device felt appropriate: the keyboard+mouse on
the desktop, the wand in the CAVE, the keyboard+mouse and the stylus
on the zSpace.

\paragraph{Loading the astronomical data}

The first task was to load the data in Unity: the software unsurprisingly
does not know about the FITS format that is commonly used in astrophysics,
so the data has to be converted. Even though this step may be trivial
for computer scientists, it is perceived as a bottleneck by many astronomers.
Rather than writing our own parser, we read the FITS file with Python
using the \texttt{astropy.io.fits} module,\footnote{Astropy is a community-developed core Python package for Astronomy
\citeyearpar{Astropy-Collaboration2013a}.} and convert it to a raw binary file that can easily be read from
any language, such as C\# used by Unity. This makes an extra step
in the visualization process, although this has to be done only once
per data cube. 

Inside Unity, the data is loaded in memory as a \emph{3D texture}.
Textures are common tools of 3D designers, although 2D textures are
far more common. 2D textures are used to draw the surface of 3D objects
to give them a more realistic look. %
To handle three-dimensional data, a possible approach is to generate
an atlas of 2D textures (see \citealt{Taylor2015b} for an application
to astronomy using the Blender rendering software); we find it simpler
to use a single 3D texture. For our purpose, we are essentially using
the texture as a look-up table, that gives the emissivity value as
a function of the three coordinates. The 3D texture has to be defined
in code, but can be saved as a Unity \emph{asset} and so be re-used.
Although floating-point texture formats exist, this is not supported
by Unity for 3D textures, so we downsize the data to an 8-bit format
\textendash{} this resolution is sufficient for the purpose of visualization,
although we note that to enable quantitative data extraction some
link will need to be kept with the original data. The texture can
be assigned as a property of a \emph{shader}, a specialized program
that runs on the GPU in parallel and that ``paints'' the pixels
on the screen.\footnote{In Unity shaders are created using the ShaderLab syntax, and are essentially
\emph{wrappers} around code snippets in standard HLSL/Cg shading language.
They are compiled to whatever target is appropriate in the current
environment (e.g. DirectX on Windows or OpenGL on macOS).} The shader itself, in the Unity terminology, gets assigned to a \emph{material}
that gets applied to an \emph{object} in the scene. 

\paragraph{Volume rendering: ray casting in the data cube}

Since we want to see the entire data cube, we are performing volume
rendering. To do this, we use the most direct method, of ray casting:
for each pixel to be rendered on the screen, a ray is cast along the
current line of sight, and along this ray the data is retrieved at
regularly spaced intervals. The values are accumulated along the line
of sight using the standard radiative transfer approximation: the
value at a point (understood here as a voxel) is interpreted as both
an emissivity (added to the \texttt{R}, \texttt{G}, \texttt{B} channels)
and an opacity (using the \texttt{alpha} channel to handle transparency).
A~key point here is that the data appears to be ``glowing'' on
its own \textendash{} there is no ambient lighting. This is useful
because we are looking at a loosely defined object amongst the background
noise, not at a pre-defined geometrical shape. 

Volume rendering is notoriously demanding in terms of processing power.
For maximum efficiency, it is done on the GPU, using custom shaders.
Our first implementation was based on the algorithm presented by \citet{Kruger2003a}
and re-used parts of an existing Unity demo project by Brian Su.\footnote{\url{https://community.unity.com/t5/Life-of-a-Unity-Game/3D-Volume-Rendering-using-Raymarching-Demo/td-p/846397}\\
and \url{https://github.com/brianasu/unity-ray-marching/tree/volumetric-textures}} This algorithm offers a straightforward way to compute all the ray
directions, but it requires three shader passes, which in Unity requires
the use of \emph{render textures}, that apply to the entire screen
rather than a single object, and this did not carry well to the stereoscopic
mode. Our second, current implementation uses a single shader adapted
from a demo by NVIDIA.\footnote{We used the ``Render to 3D Texture'' code sample from the OpenGL
SDK 10, available at \url{http://developer.download.nvidia.com/SDK/10/opengl/samples.html}.} A~demo of our approach is publicly available.\footnote{\url{https://github.com/gillesferrand/Unity-RayTracing}.}
So the data cube really is just a simple cube in the scene, that gets
``filled'' by the shader using the 3D texture. The texture is \emph{sampled}
at the location of any voxel where a data value is needed. Note that
the ``cube'' can actually be of any size, and of any aspect ratio
(we commonly choose 1:1:1 for better visibility), whatever interpolation
is needed in the texture will be done automatically on the GPU.

The display needs to be refreshed continuously as the cube is being
manipulated, with our current hardware (that is a few years old) we
obtain usable frame rates as long as we limit the number of iterations
on the GPU (steps along the ray) to about a hundred \textendash{}
so that the data cube is over-sampled along the velocity dimension
but under-sampled along the spatial dimensions.

Other techniques are possible for performing volume rendering, that
allow for a better performance at the expense of quality, such as
stacking planes, where one draws a set of planes that stay perpendicular
to the current view and sample the data on each, or \emph{splatting}
voxels, where one cuts the data cube into textured polygons that are
projected onto the screen. We leave the investigation of these optimizations
for future work.

\section{Perspectives}

The general aim of this exploratory project is not to produce \textendash yet
another\textendash{} visualization package, but to get a workbench
that allows us to experiment with the aspects that we feel are important
for our data, in particular getting a proper colouring scheme, and
overlaying other data. 

\paragraph{Colour transfer functions}

When doing volume rendering, the part that makes the data reveal itself
is the \emph{colour transfer function}, that defines the colour of
any data point (a voxel in 3D) as a function of parameters such as
the intensity of the emission at this point, or the local velocity
or spatial coordinates, or combination thereof. The colour can be
computed using a formula or looked up in a table (that can again be
stored as a texture in memory). For performance reasons it should
be specified in machine space (RGB), although it would be more sensible
to define it in perceptual space, in terms of lightness value, hue,
and chroma (see e.g. \citealp{Wijffelaars2008a,Zeileis2009a}). Note
that the colours actually seen will be altered by the integration
along the line of sight, depending on the user's viewpoint. Getting
the colour palette right for a given visualization can be a delicate
task, and we want the user to be able to dynamically adjust the mappings. 

The most straightforward thing to do, always recommended for a start,
is to use a grayscale to show the emission intensity. Mapping the
intensity value to colours may not always add much to the visualization;
defining physically-motivated transfer functions will often require
a simultaneous display of the data histogram. We find it more interesting
to use colour (and specifically the hue) to code the coordinates,
in particular the velocity coordinate since it is different from the
other two (spatial) coordinates. The most obvious choice is to use
a diverging blue vs. red palette to show the blue-shifted vs. red-shifted
parts of the galaxy. However it is all too easy to pick a red that
will pop out in the eye of the viewer, creating a perception opposite
to what is physically happening. It is possible to tweak the colours
so that ``blue comes forward'' and ``red goes back'' as it should,
by adjusting the relevant colour contrasts. One of the authors, Prof.
Jayanne English, uses such visual art techniques to clarify and support
the information (see e.g. \citealt{English2003a} for an application). 

In immersive 3D, this discussion takes a whole new turn, because we
immediately and unavoidably perceive different parts of the data cube
at different depths \textendash{} it is actually difficult not to
think of the 3D shape displayed as the actual spatial shape of the
galaxy! We want to investigate in more detail the most relevant use
of colour in different environments, from 2D to 3D to VR. 

\paragraph{Natural interaction with a data cube}

Visualizing a data cube in immersive 3D opens new possibilities, but
also creates new challenges in terms of user interaction. 

A first example is the use of a 3D cursor to make selections. On the
desktop this is always cumbersome, and commonly requires multiple
steps and adjustments of the view to get the location right. In VR,
one can use an actual 3D pointer, that can be one's finger, or a tool
like a wand or stylus, as long as it is being tracked inside the volumetric
display. This has many possible applications, like: display the coordinates
and data value of a point, select all the points having the same value
(live iso-contouring), or show where this point falls on the histogram
(using multiple linked views).

Another useful example is to overlay other data. Since at other wavelengths
(such as optical) data are obtained in the form of images, there is
a need to overlay 2D data on top of the 3D data, as floating panels.
One could then step the image through the cube (along the velocity
axis) to find matching features, and segment the image and attach
the segments where such matching features are seen, in order to disentangle
the different parts of the galaxy. This can be done on the desktop
by stepping through the velocity channels, but in real 3D one can
easily look at any angle, and see for instance contiguous features
along the velocity axis (without having to remember and reconstruct
them mentally). Since radio observations may be done at different
wavelengths corresponding to different emission lines (e.g. from HI
and CO), another possibility is to superpose two (or more) 3D data
cubes. This is also relevant for optical/UV data obtained with Integral
Field Units, for other objects like planetary nebulae or supernova
remnants. The data would be merged voxel by voxel, just like when
creating 2D image ``composites'' from data obtained with different
filters or instruments, only in a more complex way because of the
view-dependent aspect. 

\begin{figure*}
\noindent \centering{}\includegraphics[height=0.4\textheight]{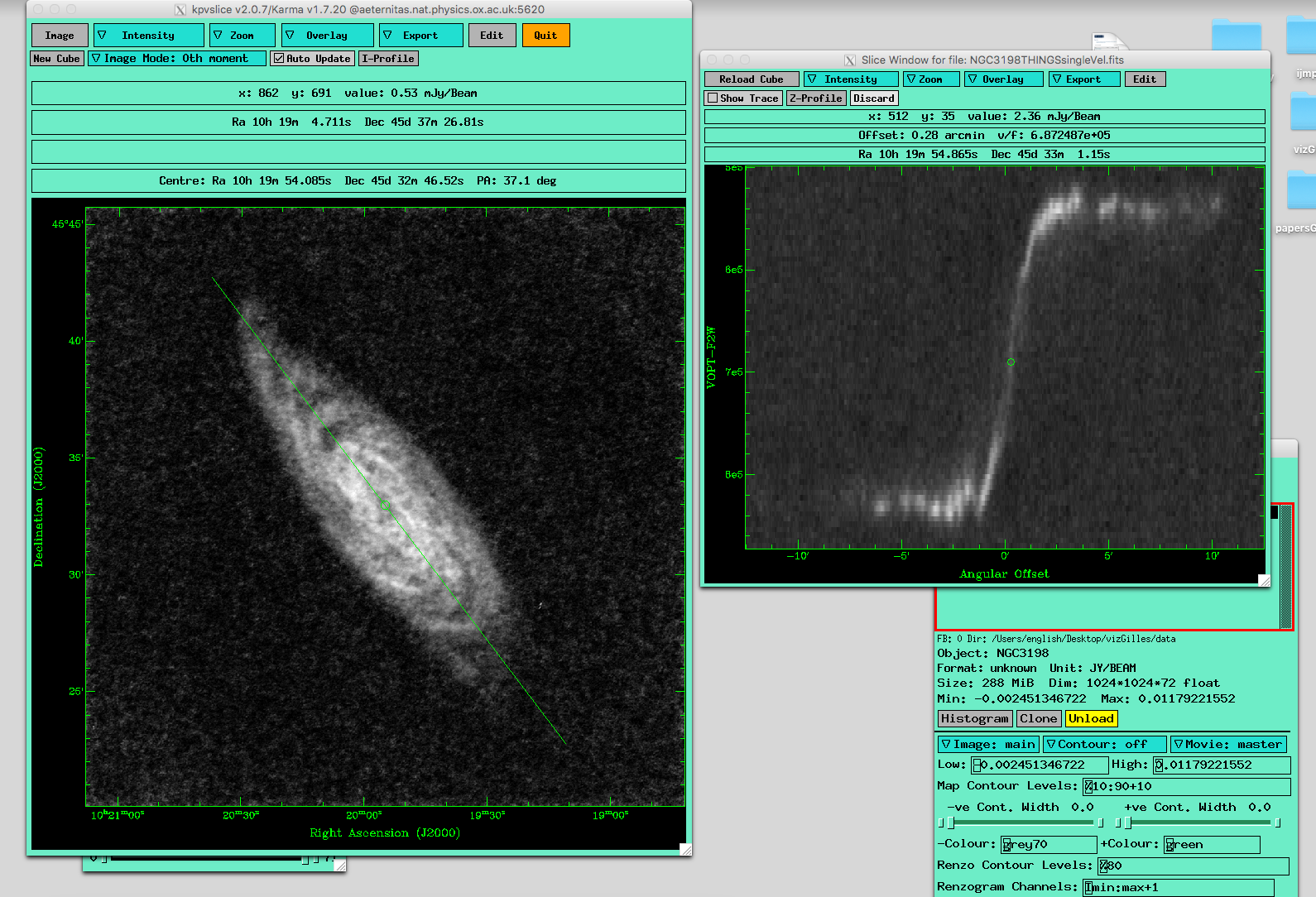}\caption{\label{fig:Karma}Screenshot of a typical radio-astronomer's desktop
on a personal computer. The program \texttt{kpvslice} from the Karma
software suite is being used. The window on the left shows the moment~0
map of the data cube, that is a projection along the velocity channels,
where one can see the galaxy's physical outline. The window on the
right shows a position-velocity diagram made along the spatial axis
shown as the green line on left plot, where one can see the rotation
curve of the galaxy (uncorrected for its inclination to the plane
of the sky).}
\end{figure*}

\begin{figure*}
\noindent \centering{}\includegraphics[height=0.4\textheight]{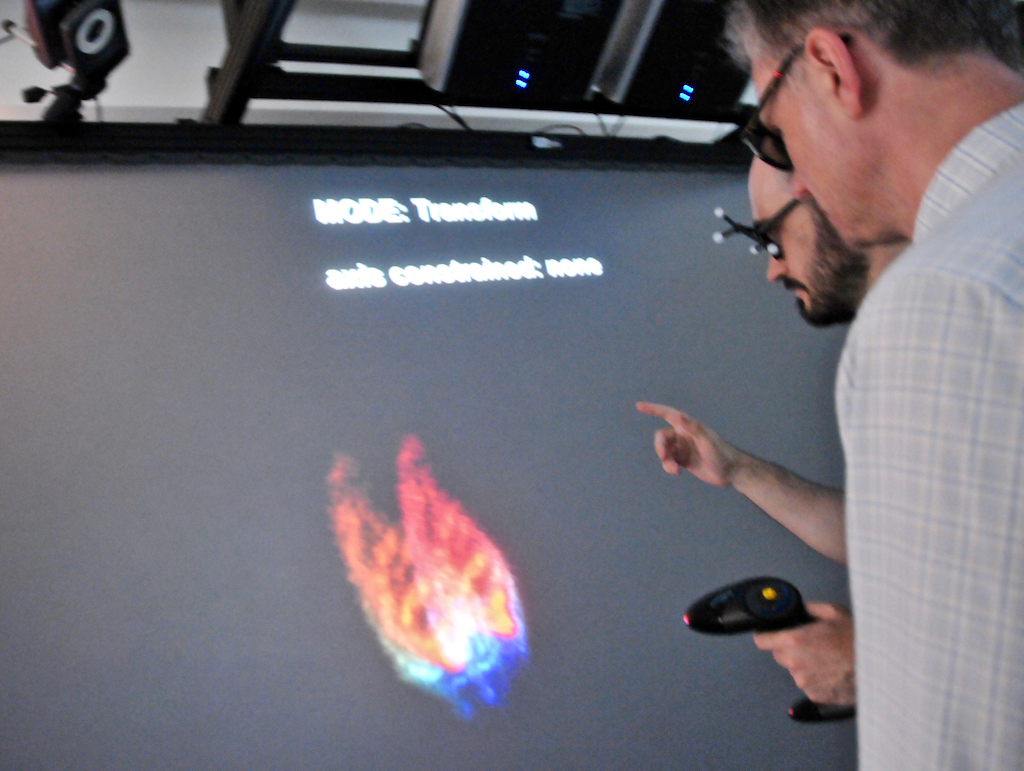}\caption{\label{fig:visitorCAVE}Photograph showing the lead author inspecting
the galaxy data cube with a visitor in the CAVE. The two images visible
on the screen are merged into a 3D view thanks to the polarized glasses.
Also note the reflectors that allow to track the gaze of the person
looking at the data, and the wand used to interact with it. For the
two persons the data cube appears as a 3D cloud floating in the air,
the pointed finger is actually ``touching'' the edge of the galaxy
\textendash{} no photograph can convey the actual experience.}
\end{figure*}

Comparing figures~\ref{fig:Karma} and~\ref{fig:visitorCAVE}, it
is apparent that our current prototype is in the realm of qualitative
rather than quantitative analysis. To make it a useful tool for science,
we want to eventually include all the steps of the discovery process:
explore, explain, extract. One of the authors, Prof. Pourang Irani,
is devising new ways of interacting with data (e.g. \citealt{Ens2014a}),
and we plan to integrate the knowledge of the Human-Computer Interaction
field to produce intuitive visualization tools and techniques that
people will actually want to use for their research.

\paragraph{Conclusion and an invitation}

The first feedback that we have gathered, amongst members of our astronomy
group at the University of Manitoba, and during the 2016 Annual General
Meeting of the Canadian Astronomical Society (CASCA), was positive.\footnote{This paper serves as the proceedings of the presentation that the
first author gave at this meeting, the slides are available at \url{http://www.physics.umanitoba.ca/~gferrand/docs/FERRAND_2016-06-01_CASCA-talk.pdf}.} Most people were impressed by the technology, in part because nearly
all of them were experiencing it for the first time \textendash{}
with this work we want to raise awareness about 3D displays, amongst
radio astronomers in particular and in the astronomy community at
large. We hope to build a special interest group, to keep moving forward
and share experiences about best practices as well as caveats in this
new, exciting area of scientific visualization. \\
~

\acknowledgements{}

This pilot project was funded at the University of Manitoba by the
Faculty of Science's Interdisciplinary/New Directions Research Collaboration
Initiation Grants and by the University Collaborative Research Program
(UCRP). G.F. warmly thanks the HCI lab team for hosting him during
this work. 

\bibliographystyle{plainnat}
\bibliography{visu3D}

\begin{thebibliography}{15}
\providecommand{\natexlab}[1]{#1}
\providecommand{\url}[1]{\texttt{#1}}
\expandafter\ifx\csname urlstyle\endcsname\relax
  \providecommand{\doi}[1]{doi: #1}\else
  \providecommand{\doi}{doi: \begingroup \urlstyle{rm}\Url}\fi

\bibitem[Barnes et~al.(2006)Barnes, Fluke, Bourke, and Parry]{Barnes2006a}
D.~G. Barnes, C.~J. Fluke, P.~D. Bourke, and O.~T. Parry.
\newblock An advanced, three-dimensional plotting library for astronomy.
\newblock \emph{\pasa}, 23:\penalty0 82--93, July 2006.
\newblock \doi{10.1071/AS06009}.
\newblock URL \url{http://dx.doi.org/10.1071/AS06009}.

\bibitem[Cioc et~al.(2013)Cioc, Djorgovski, Donalek, Lawler, Sauer, and
  Longo]{Cioc2013a}
A.~Cioc, S.~G. Djorgovski, C.~Donalek, E.~Lawler, F.~Sauer, and G.~Longo.
\newblock Data visualization using immersive virtual reality tools.
\newblock In \emph{American Astronomical Society Meeting Abstracts \#221},
  volume 221 of \emph{American Astronomical Society Meeting Abstracts}, page
  240.20, January 2013.
\newblock URL \url{http://adsabs.harvard.edu/abs/2013AAS...22124020C}.

\bibitem[Collaboration et~al.(2013)Collaboration, Robitaille, Tollerud,
  Greenfield, Droettboom, Bray, Aldcroft, Davis, Ginsburg, Price-Whelan,
  Kerzendorf, Conley, Crighton, Barbary, Muna, Ferguson, Grollier, Parikh,
  Nair, Unther, Deil, Woillez, Conseil, Kramer, Turner, Singer, Fox, Weaver,
  Zabalza, Edwards, Azalee~Bostroem, Burke, Casey, Crawford, Dencheva, Ely,
  Jenness, Labrie, Lim, Pierfederici, Pontzen, Ptak, Refsdal, Servillat, and
  Streicher]{Astropy-Collaboration2013a}
Astropy Collaboration, T.~P. Robitaille, E.~J. Tollerud, P.~Greenfield,
  M.~Droettboom, E.~Bray, T.~Aldcroft, M.~Davis, A.~Ginsburg, A.~M.
  Price-Whelan, W.~E. Kerzendorf, A.~Conley, N.~Crighton, K.~Barbary, D.~Muna,
  H.~Ferguson, F.~Grollier, M.~M. Parikh, P.~H. Nair, H.~M. Unther, C.~Deil,
  J.~Woillez, S.~Conseil, R.~Kramer, J.~E.~H. Turner, L.~Singer, R.~Fox, B.~A.
  Weaver, V.~Zabalza, Z.~I. Edwards, K.~Azalee~Bostroem, D.~J. Burke, A.~R.
  Casey, S.~M. Crawford, N.~Dencheva, J.~Ely, T.~Jenness, K.~Labrie, P.~L. Lim,
  F.~Pierfederici, A.~Pontzen, A.~Ptak, B.~Refsdal, M.~Servillat, and
  O.~Streicher.
\newblock Astropy: A community python package for astronomy.
\newblock \emph{\aap}, 558:\penalty0 A33, October 2013.
\newblock \doi{10.1051/0004-6361/201322068}.
\newblock URL \url{http://dx.doi.org/10.1051/0004-6361/201322068}.

\bibitem[English et~al.(2003)English, Norris, Freeman, and Booth]{English2003a}
J.~English, R.~P. Norris, K.~C. Freeman, and R.~S. Booth.
\newblock Ngc 3256: Kinematic anatomy of a merger.
\newblock \emph{\aj}, 125:\penalty0 1134--1149, March 2003.
\newblock \doi{10.1086/367914}.
\newblock URL \url{http://dx.doi.org/10.1086/367914}.

\bibitem[Ens et~al.(2014)Ens, Hincapi{\'e}-Ramos, and Irani]{Ens2014a}
Barrett Ens, Juan~David Hincapi{\'e}-Ramos, and Pourang Irani.
\newblock Ethereal planes: A design framework for 2d information space in 3d
  mixed reality environments.
\newblock In \emph{Proceedings of the 2Nd ACM Symposium on Spatial User
  Interaction}, SUI '14, pages 2--12, New York, NY, USA, 2014. ACM.
\newblock ISBN 978-1-4503-2820-3.
\newblock \doi{10.1145/2659766.2659769}.
\newblock URL \url{http://doi.acm.org/10.1145/2659766.2659769}.

\bibitem[Fluke and Barnes(2016)]{Fluke2016a}
C.~J. Fluke and D.~G. Barnes.
\newblock The ultimate display.
\newblock \emph{ArXiv e-prints}, January 2016.
\newblock URL \url{http://adsabs.harvard.edu/abs/2016arXiv160103459F}.

\bibitem[Hassan and Fluke(2011)]{Hassan2011a}
A.~Hassan and C.~J. Fluke.
\newblock Scientific visualization in astronomy: Towards the petascale
  astronomy era.
\newblock \emph{\pasa}, 28:\penalty0 150--170, June 2011.
\newblock \doi{10.1071/AS10031}.
\newblock URL \url{http://dx.doi.org/10.1071/AS10031}.

\bibitem[Hassan et~al.(2011)Hassan, Fluke, and Barnes]{Hassan2011b}
A.~H. Hassan, C.~J. Fluke, and D.~G. Barnes.
\newblock Interactive visualization of the largest radioastronomy cubes.
\newblock \emph{\na}, 16:\penalty0 100--109, February 2011.
\newblock \doi{10.1016/j.newast.2010.07.009}.
\newblock URL \url{http://dx.doi.org/10.1016/j.newast.2010.07.009}.

\bibitem[Kruger and Westermann(2003)]{Kruger2003a}
J.~Kruger and R.~Westermann.
\newblock Acceleration techniques for gpu-based volume rendering.
\newblock In \emph{Visualization, 2003. VIS 2003. IEEE}, pages 287--292, Oct
  2003.
\newblock \doi{10.1109/VISUAL.2003.1250384}.
\newblock URL \url{http://dx.doi.org/10.1109/VISUAL.2003.1250384}.

\bibitem[Norris(1994)]{Norris1994a}
R.~P. Norris.
\newblock The challenge of astronomical visualisation.
\newblock In D.~R. Crabtree, R.~J. Hanisch, and J.~Barnes, editors,
  \emph{Astronomical Data Analysis Software and Systems III}, volume~61 of
  \emph{Astronomical Society of the Pacific Conference Series}, page~51, 1994.
\newblock URL \url{http://adsabs.harvard.edu/abs/1994ASPC...61...51N}.

\bibitem[Punzo et~al.(2015)Punzo, van~der Hulst, Roerdink, Oosterloo,
  Ramatsoku, and Verheijen]{Punzo2015a}
D.~Punzo, J.~M. van~der Hulst, J.~B.~T.~M. Roerdink, T.~A. Oosterloo,
  M.~Ramatsoku, and M.~A.~W. Verheijen.
\newblock The role of 3-d interactive visualization in blind surveys of hi in
  galaxies.
\newblock \emph{Astronomy and Computing}, 12, September 2015.
\newblock URL \url{http://adsabs.harvard.edu/abs/2015arXiv150506976P}.

\bibitem[Rosolowsky et~al.(2015)Rosolowsky, Kern, Federl, Jacobs, Loveland,
  Taylor, Sivakoff, and Taylor]{Rosolowsky2015b}
E.~Rosolowsky, J.~Kern, P.~Federl, J.~Jacobs, S.~Loveland, J.~Taylor,
  G.~Sivakoff, and R.~Taylor.
\newblock The cube analysis and rendering tool for astronomy.
\newblock In A.~R. Taylor and E.~Rosolowsky, editors, \emph{Astronomical Data
  Analysis Software an Systems XXIV (ADASS XXIV)}, volume 495 of
  \emph{Astronomical Society of the Pacific Conference Series}, page 121,
  September 2015.
\newblock URL
  \url{http://aspbooks.org/custom/publications/paper/495-0121.html}.

\bibitem[Taylor(2015)]{Taylor2015b}
R.~Taylor.
\newblock Frelled: A realtime volumetric data viewer for astronomers.
\newblock \emph{Astronomy and Computing}, 13:\penalty0 67--79, November 2015.
\newblock \doi{10.1016/j.ascom.2015.10.002}.
\newblock URL \url{http://dx.doi.org/10.1016/j.ascom.2015.10.002}.

\bibitem[Wijffelaars et~al.(2008)Wijffelaars, Vliegen, Van~Wijk, and Van
  Der~Linden]{Wijffelaars2008a}
Martijn Wijffelaars, Roel Vliegen, Jarke~J. Van~Wijk, and Erik-Jan Van
  Der~Linden.
\newblock Generating color palettes using intuitive parameters.
\newblock \emph{Computer Graphics Forum}, 27\penalty0 (3):\penalty0 743--750,
  2008.
\newblock ISSN 1467-8659.
\newblock \doi{10.1111/j.1467-8659.2008.01203.x}.
\newblock URL \url{http://dx.doi.org/10.1111/j.1467-8659.2008.01203.x}.

\bibitem[Zeileis et~al.(2009)Zeileis, Hornik, and Murrell]{Zeileis2009a}
Achim Zeileis, Kurt Hornik, and Paul Murrell.
\newblock Escaping rgbland: Selecting colors for statistical graphics.
\newblock \emph{Computational Statistics \& Data Analysis}, 53:\penalty0
  3259--3270, 2009.
\newblock URL \url{http://epub.wu.ac.at/1692/}.

\end{thebibliography}

\clearpage{}


\end{document}